\documentclass[twocolumn,secnumarabic,amssymb, nobibnotes, aps, prd]{revtex4-1}
\usepackage{graphicx}
\usepackage{amsmath}
\usepackage{hyperref}
\usepackage{aasmacros}

\usepackage{color}

\begin{document}

\title{Cosmological constraints from noisy convergence maps through deep learning}

\author{Janis Fluri}%
\email{janis.fluri@phys.ethz.ch}
\author{Tomasz Kacprzak}
\author{Alexandre Refregier}
\author{Adam Amara}
\affiliation{Institute of Particle Physics and Astrophysics, Department of Physics, ETH Zurich, Switzerland}
\author{Aurelien Lucchi}
\author{Thomas Hofmann}
\affiliation{Data Analytics Lab, Department of Computer Science, ETH Zurich, Switzerland}
\date{\today}%

\begin{abstract}
Deep learning is a powerful analysis technique that has recently been proposed as a method to constrain cosmological parameters from weak lensing mass maps.
Due to its ability to learn relevant features from the data, it is able to extract more information from the mass maps than the commonly used power spectrum, and thus achieve better precision for cosmological parameter measurement.
We explore the advantage of Convolutional Neural Networks (CNN) over the power spectrum for varying levels of shape noise and different smoothing scales applied to the maps.
We compare the cosmological constraints from the two methods in the $\Omega_M-\sigma_8$ plane for sets of 400 deg$^2$ convergence maps.
We find that, for a shape noise level corresponding to 8.53 galaxies/arcmin$^2$ and the smoothing scale of $\sigma_s = 2.34$ arcmin, the network is able to generate 45\% tighter constraints.
For smaller smoothing scale of $\sigma_s = 1.17$ the improvement can reach $\sim 50 \%$, while for larger smoothing scale of $\sigma_s = 5.85$, the improvement decreases to 19\%.
The advantage generally decreases when the noise level and smoothing scales increase.
We present a new training strategy to train the neural network with noisy data, as well as considerations for practical applications of the deep learning approach.

\end{abstract}

\maketitle

\section{Introduction}

Weak gravitational lensing (WL) is a powerful probe to examine the large scale structure of the universe (see e.g. \cite{Bartelmann2001} for a review). By measuring the tiny distortions in the shapes of faint background galaxies it is possible to reconstruct the projected matter density known as convergence. Many WL surveys, such as the Canada-France-Hawaii Telescope Lensing Survey (CFHTLenS\footnote{\url{cfhtlens.org}}) \cite{Heymans2013}, the Kilo-Degree Survey (KiDS\footnote{\url{kids.strw.leidenuniv.nl}}) \cite{Hildebrandt2016} or the Dark Energy Survey (DES\footnote{\url{darkenergysurvey.org}}) \cite{Troxel2016}, have successfully demonstrated the capabilities of this method to constrain cosmological parameters and the quality and quantity of the collected data is expected to increase further in the near future.

Common ways to analyze the data collected by these surveys include the two and three point correlation function (e.g. \cite{Hildebrandt2016, Troxel2016, Semboloni2011, Takada2003, Fu2014}) and weak lensing peak statistics (e.g. \cite{Kacprzak2016a, Marian2012, Marian2013, Liu2016a, Shan2017, Liu2014}), among others. These techniques have been used separately to constrain cosmological parameters and they have also been combined which usually improves the constraints (e.g. \cite{Liu2015b, Peel2016, Fu2014}) since they use different information extracted from weak lensing maps.

Even though these summary statistics are highly successful at constraining cosmological parameters, there is still an ongoing search for new techniques with other advantages. A promising approach is to analyze weak lensing maps using deep learning, a machine learning (ML) technique that automatically extract features from data, as opposed to using hand-designed summary statistics. The potential of such techniques to constrain cosmological parameters has recently been demonstrated in \cite{Schmelzle2017, Gupta2018}.

ML has also been applied to other astrophysical problems. Examples include the mitigation of radio frequency interference \cite{Akeret2016}, the use of generative adversarial networks to simulate observational data \cite{Mustafa2017, Rodriguez2018}, populating dark matter only simulations with baryonic galaxies \cite{Agarwal2017}, fast point spread function modeling \cite{Herbel2018Psf} and strong gravitational lensing \cite{Levasseur2017}.

Convolutional neural networks (CNN) have been shown to be able to successfully discriminate weak lensing maps from different cosmologies \cite{Schmelzle2017}. The CNN used in \cite{Schmelzle2017} was a classification network and was able to distinguish different cosmologies even when the input maps contained a realistic amount of shape noise. CNNs were also used by \cite{Gupta2018} to study the information content of noise free, high resolution convergence maps. They used a CNN as a regression network and were able to constrain cosmological parameters but they did not address the problem of using noisy convergence maps.

In this work we examine the constraining power of CNNs compared to a standard power spectrum analysis for varying levels of noise and smoothing scales applied to the convergence maps. We explore different survey regimes by adding different levels of shape noise, corresponding to current and future weak lensing surveys. Gaussian smoothing corresponds to the quality and realism of the numerical simulations that are used to train the CNN.
Gaussian smoothing kernels with a small width $\sigma_s$ leave a lot of non-Gaussian information in the small scale structures, allowing the network to extract more information and to converge faster. However, these small scale structures are very hard to model. Many effects can influence the small scale structure, including baryonic effects, which  start to have an impact on the power spectrum at $\ell \sim 4000$ or even lower \cite{Osato2015baryonic,Mohammed2014,McCarthy2018bahamas}, the simulations resolution effects related to the number of particles, and the simulations parameters, such as the softening scale. Finally, N-body prediction codes start to slightly disagree at $k \sim 1 h$ Mpc$^{-1}$ \cite{Schneider2016challenge}.

We consider a 400 deg$^2$ survey, similar to KiDS-450 \cite{Hildebrandt2016}, consisting of 4 independent patches of 100 deg$^2$ noisy convergence maps. We produce the necessary data by generating a large amount of N-body simulations using the fast \texttt{L-PICOLA} code \cite{Howlett2015}. Using these simulations we then generate convergence maps using the fast convergence maps generator \texttt{Ufalcon} \cite{Sgier2018}. Afterwards we train a CNN to predict the total matter density $\Omega_M$ and the fluctuation amplitude $\sigma_8$ using these convergence maps as inputs. To have a point of reference we compare the cosmological constraints generated from the CNN to constraints generated from a power spectrum analysis on the same maps.

The noise we add to our data corresponds to Gaussian random noise, which we used to model shape noise \cite{Pires2009, Refregier2012}. We consider different levels of shape noise including realistic noise levels expected from ground- or space-based observations. In this work we did not include any other effects or systematics, such as the magnification bias that is caused by preferential selection of magnified source galaxies \cite{Liu2013a}, the intrinsic alignment of galaxies around large clusters \cite{Heavens2000}, photometric redshift error \cite{Bonnett2016, Cavuoti2017} and the shear calibration bias \cite{Zuntz2017, FenechConti2017}. We leave the treatment of these effects to future work.

This paper is structured in the following way. In section \ref{sec:data} we describe how we generated our data to train the CNN. The power spectrum analysis is explained in section \ref{sec:power}. The setup and training of the CNN is covered in section \ref{sec:CNN}. Afterwards we present our results in section \ref{sec:results}, which is followed by a conclusion and outlook in section \ref{sec:conc}. In appendix \ref{ap:simulations} we give further insights into our N-body simulation settings, appendix \ref{ap:convmaps} gives an overview of the convergence map generator \texttt{Ufalcon} \cite{Sgier2018}. The effect of a super-sample covariance matrix is examined in appendix \ref{ap:supersample}. Potential overfitting of the CNN and the training and validation loss of the network are discussed in appendix \ref{ap:loss}. In appendix \ref{ap:cov} we show the effect of our covariance matrix interpolation scheme. In appendix \ref{ap:robust} we analyze the cosmological constraints of our analysis depending on the mock observation.

\section{Data Generation \label{sec:data}}

\begin{figure}
\includegraphics[width=0.5\textwidth]{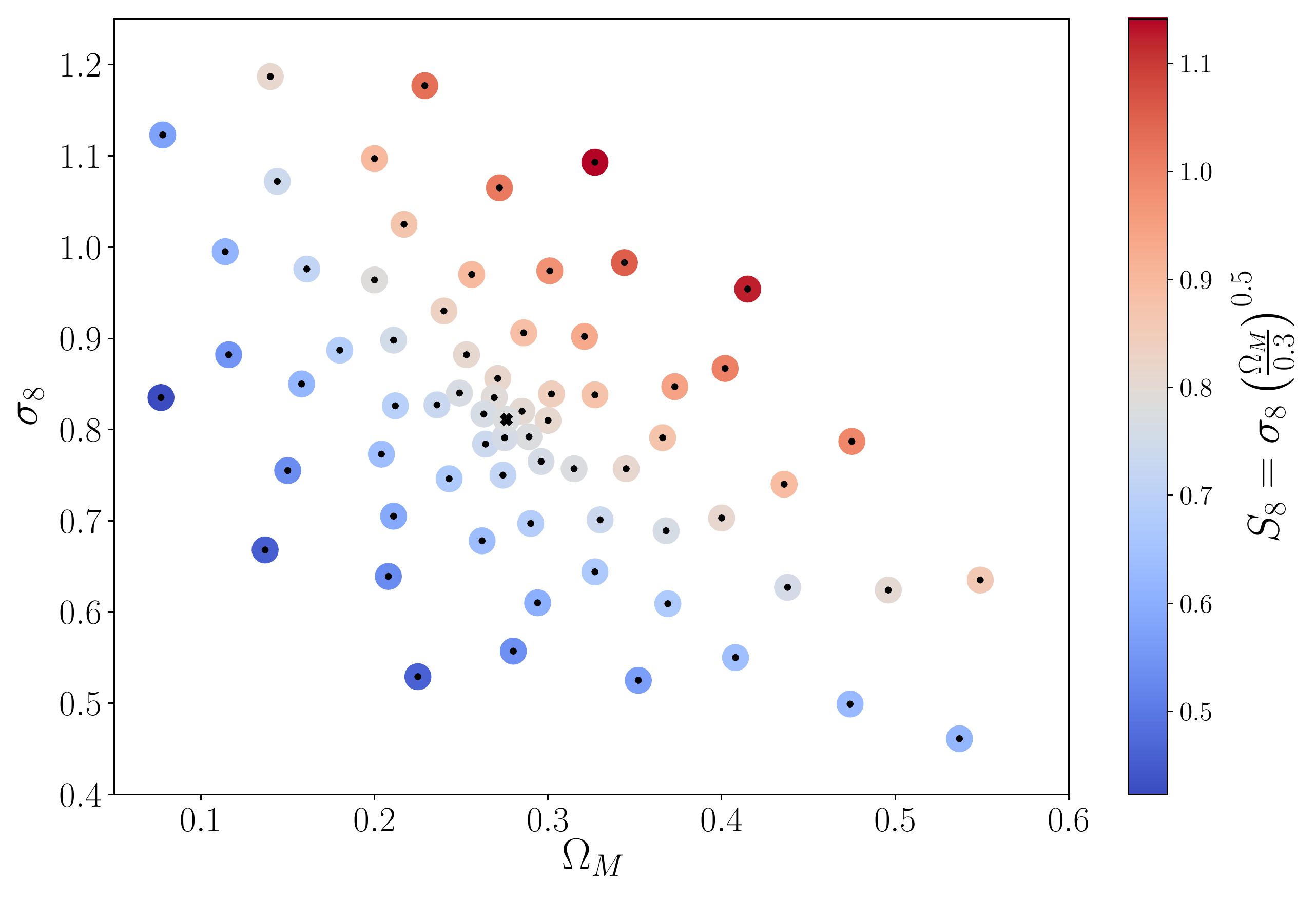}
\caption{Grid of the simulated cosmologies. The black cross in the middle corresponds to our fiducial cosmology. The color is given by the degeneracy parameter $S_8$. \label{fig:simugrid}}
\end{figure}

To generate the convergence maps necessary to train the CNN we produced a large amount of N-body simulations using the fast \texttt{L-PICOLA} code  \cite{Howlett2015}. \texttt{L-PICOLA} is a particle mesh code that relies on the comoving Lagrangian acceleration (COLA) method to produce accurate simulations orders of magnitudes faster than comparable full N-body simulations \cite{Howlett2015, Sgier2018}. The simulations were produced in almost the same way as in \cite{Fluri2018}. A detailed description of the simulation settings can be found in appendix \ref{ap:simulations}. Assuming a flat $\Lambda$CDM universe we simulated a total of 70 different cosmologies. The parameters of our fiducial cosmology were set to $\Omega_M = 0.276$, $\sigma_8=0.811$, $h = 0.7$, $n_s = 0.961$ and $\Omega = 0.045$. All other cosmologies were generated by only changing $\Omega_M$ and $\sigma_8$ (and $\Omega_\Lambda$ such that the universe remains flat). The whole simulation grid is shown in figure \ref{fig:simugrid}. All points of the simulation grid lay on ellipses $\mathcal{E}(t) = (a\cos(t), b\sin(t))$ centered around our fiducial cosmologies and for each ellipse we chose equally spaced values of $t \in [0, 2\pi)$. The density of the ellipses increases stepwise towards the fiducial cosmology and we only considered points inside our prior range of $\Omega_m \in [0, 0.7]$ and $\sigma_8 \in [0.4, 1.4]$.
We generated a total of 40 simulations for the fiducial cosmology and 4 simulations for the remaining cosmologies on the grid. The simulations were run in snapshot mode and snapshots were outputted from $z=1.5$ to $z=0.1$ every $\Delta z = 0.01$.  For each simulation we used a nested box approach to get a higher particle resolution at lower redshifts. One simulation therefore consisted of two independently evolved boxes. The first box had a size of 6 Gpc and was used for the redshift range $0.1 \leq z < 0.8$ and the second box had a size of 9 Gpc and was used for the redshift range $0.8 \leq z  \leq 1.5$. To further reduce the computation time of these simulations we used a replication scheme and built these large boxes out of multiple smaller ones (see Appendix \ref{ap:simulations}).

One disadvantage of \texttt{L-PICOLA} is that it does not resolve the small scale structures as accurately as full TreePM codes like Gadget-2\citep{Springel2005}. The simulations used in this work are agreeing with theoretical predictions of the power spectrum up to $\ell\sim1000$ (see figure \ref{fig:power_spec}) and are therefore not fully realistic. However, we expect that the CNN and the power spectrum analysis are affected in the same way, such that the comparison remains fair.

\subsection{Convergence Maps}

To generate full sky convergence maps out of the N-body simulations we used the fast convergence map generator \texttt{Ufalcon} \cite{Sgier2018}. \texttt{Ufalcon} follows the procedure described in the appendix of \cite{Teyssier2009} and uses the Hierarchical Equal Area iso-Latitude Pixelization tool \cite{Gorski2004} (\texttt{HEALPix}\footnote{http://healpix.sourceforge.net}). A detailed description can be found in \cite{Sgier2018} and a quick overview is given in appendix \ref{ap:convmaps}. Out of each snapshot we cut out a shell of particles with a thickness of $\Delta z = 0.01$ and used \texttt{Ufalcon} to generate full sky convergence maps with a
Smail et al. \cite{Smail1994} source galaxy redshift distribution of
\begin{equation}
n(z) \propto z^2 \exp\left(-\frac{z}{0.24}\right),
\end{equation}
which corresponds to a typical redshift distribution for stage-2 weak lensing experiments. All full sky maps generated this way had an \texttt{HEALPix}\cite{Gorski2004} resolution of $\mathrm{nside} = 1024$.

\begin{figure*}[t]
\includegraphics[width=1.0\textwidth]{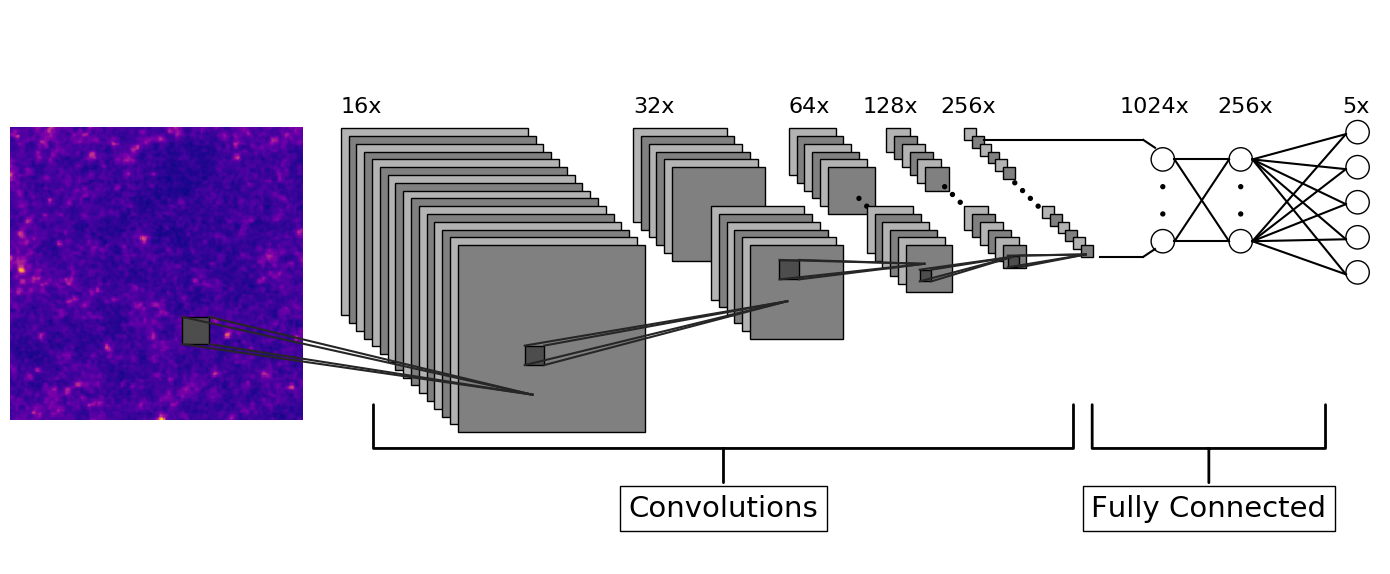}
\caption{Sketch of the CNN's architecture: 5 convolutional layers, 2 fully connected layers and one output layer. The input of the CNN is one 10x10 deg$^2$ convergence map. The covergence map shown here was smoothed with a Gaussian smoothing kernel of width $\sigma_s = 2.34$ arcmin. \label{fig:architecture}}
\end{figure*}

In order to obtain a dataset of convergence maps suitable to analyze with a typical CNN we had to project the full sky convergence maps onto a flat surface. This was done using the gnomonic projection scheme implemented in \texttt{HEALPix}\cite{Gorski2004}. We generated 768 different, flat 10 x 10 deg$^2$ patches with 256 x 256 pixels out of each full sky convergence map. This resulted in a total of 3'072 patches for each cosmology (30'720 for the fiducial cosmology). An example of such a convergence maps, smoothed with a Gaussian smoothing kernel of width $\sigma_s = 2.34$ arcmin is shown as the input layer of the CNN in figure \ref{fig:architecture}. The center of each patch corresponded to a vector pointing at the center of a \texttt{HEALPix} map of $\mathrm{nside} = 8$ and the orientation of the edges of the projection was chosen randomly. Choosing the centers of the patches at \texttt{HEALPix} pixel positions ensured an even distribution of the patches around the sphere. One should note that 768 patches of 100 deg$^2$ account for roughly 1.8 full spheres leading to overlapping patches. However, the overlapping region always had a random orientation since the orientation of the patches was chosen randomly. We chose a configuration similar to the KiDS-450 \cite{Hildebrandt2016} survey and split our dataset into groups of 4 non-overlapping patches. We considered each group as one realization of our examined region of the sky.

Further we analyzed these patches with different noise levels and applied different Gaussian smoothing kernels. To mimic observational noise we added Gaussian random noise to each pixel of the full sky convergence maps before projecting them onto a flat surface. The width of the Gaussian noise distribution was chosen as
\begin{equation}
\sigma_\mathrm{noise} = \frac{\sigma_e}{\sqrt{An}},
\end{equation}
where $A$ corresponds to the pixel area, $n$ to the galaxy number density and $\sigma_e$ is the ellipticity dispersion. We set $\sigma_e = 0.290$ which corresponds to the ellipticity dispersion of KiDS-450\cite{Hildebrandt2016}. Besides the noise free case we considered 4 different noise levels corresponding to different galaxy number densities $n \in \{100\ \mathrm{arcmin}^{-2}, 30\ \mathrm{arcmin}^{-2}, 15\ \mathrm{arcmin}^{-2}, 8.53\ \mathrm{arcmin}^{-2} \}$ where $n = 8.53\ \mathrm{arcmin}^{-2}$ is the effective galaxy number density of KiDS-450\cite{Hildebrandt2016} and $\sim 30$ $\mathrm{arcmin}^{-2}$ is the expected galaxy number density for surveys like LSST \cite{LSST2013}, Euclid \cite{Euclid2011} and HSC \footnote{http://hsc.mtk.nao.ac.jp/ssp/science/weak-lensing-cosmology/}. The other noise levels were chosen to examine how the advantage of the CNN scales with respect to increasing shape noise.

Gaussian smoothing kernels were always applied after the noise was added and on the projected maps. We considered 5 different Gaussian smoothing kernels with widths $\sigma_s \in \{0.5l_p, 1l_p, 1.5l_p, 2l_p, 2.5l_p \}$, where $l_p = 2.34$ arcmin was the length of one pixel of the projected patches.

\begin{figure}
\includegraphics[width=0.5\textwidth]{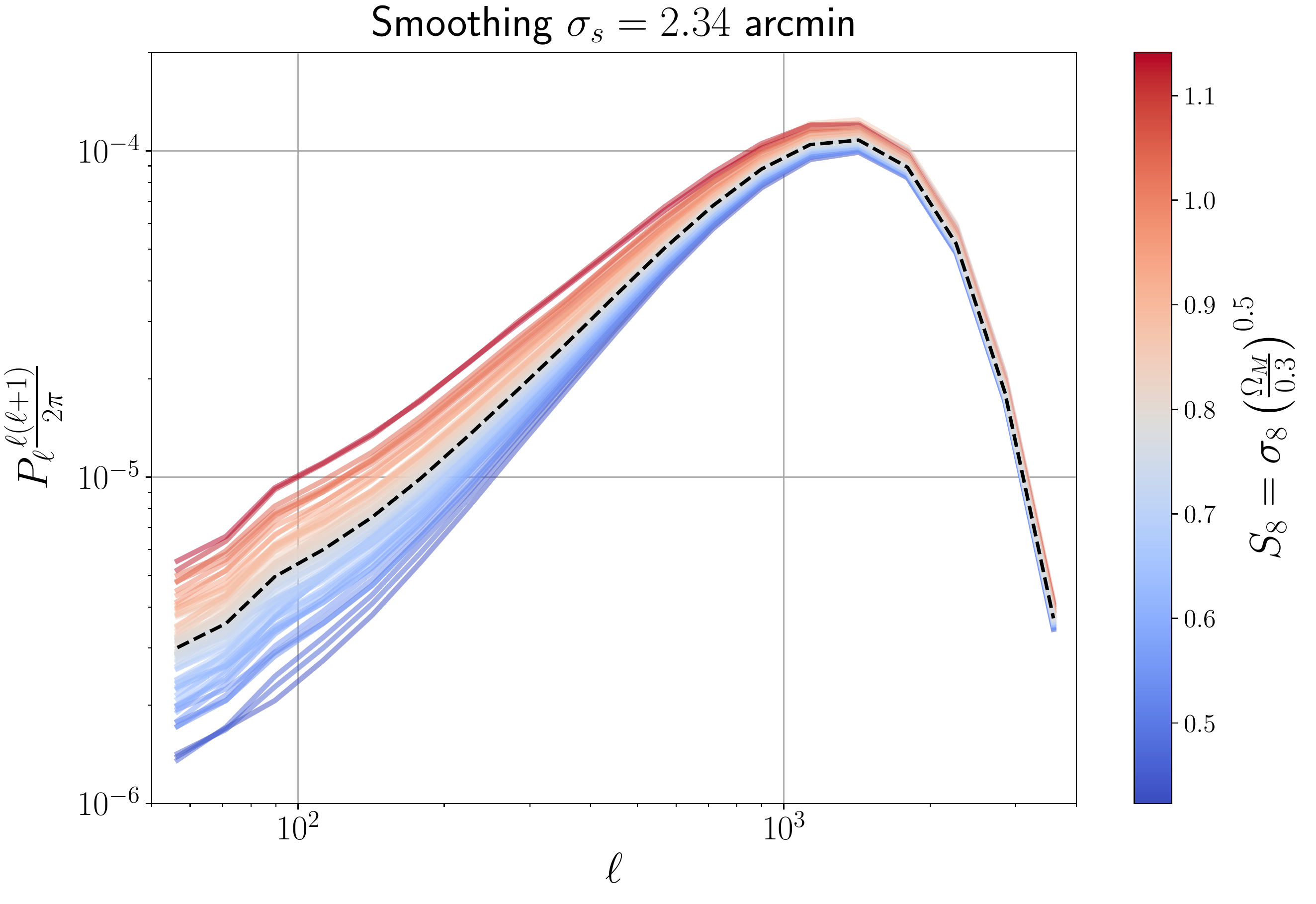}
\caption{Average power spectra of all patches of all simulated cosmologies. The black dashed line corresponds to our fiducial cosmology. The color is given by the degeneracy parameter $S_8$. We added Gaussian random noise corresponding to a galaxy number-density of $n = 8.53$ galaxies/arcmin$^2$ to the patches. The patches were smoothed with a Gaussian kernel of width $\sigma_s = 2.34$ arcmin before the spectra were calculated. \label{fig:spectra}}
\end{figure}

\section{Power Spectrum Analysis \label{sec:power}}

A power spectrum analysis was done to establish a point of reference for the performance of the CNN. All spectra were binned into 19 logarithmically spaced bins between $\ell \in [50, 4000]$. However, one should note that, depending on the applied smoothing kernel, most of the considered high $\ell$'s were smoothed out to zero (see for example figure \ref{fig:spectra}). The Gaussian smoothing was applied to have a fair comparison of the resulting cosmological constraints between the network and the power spectrum. While the power spectrum can only access information in our considered $\ell$ range, the network could potentially access information from all scales. The smoothing was therefore applied to give both methods access to the same scales and to smooth out small scales which are very hard to model theoretically.

The power spectrum was calculated using \texttt{LensTools}\cite{Petri2016}. We calculated the power spectrum for each patch and each combination of noise level and smoothing scale. The power spectra of 4 patches was then averaged to get the power spectra of our examined region.

\subsection{Inference \label{sec:powerinfer}}

We generate cosmological constraints using the likelihood proposed by \cite{Selletin2016, Selletin2017} that includes the uncertainty of the covariance matrix generated from simulations. To do thus we calculated the average power spectrum of the patches for every simulated cosmology. This was done for all combinations of noise levels and smoothing scales that we considered. An example of these mean spectra is shown in figure \ref{fig:spectra}. The mean power spectra of the simulations were then interpolated to estimate the power spectrum $\mathbf{p}(\theta)$ for any given cosmology $\theta = (\Omega_M, \sigma_8)$. The interpolation procedure was performed using smooth bivariant splines\footnote{\texttt{SmoothBivariateSpline} from the \texttt{scipy.interpolate} package}. The covariance matrix for our fiducial cosmology was then calculated in the following way
\begin{equation}
\hat{\Sigma} = \frac{1}{N_s-1}\sum_{i=1}^{N_s}\left(\hat{\mathbf{p}}_i-\bar{\mathbf{p}}\right)\left(\hat{\mathbf{p}}_i-\bar{\mathbf{p}}\right)^T, \label{eq:unbiasedcov}
\end{equation}
where $\hat{\mathbf{p}}_i$ is a simulated realization of our examined region (average over 4 patches), $N_s = 7680$ is the total number of such realizations and $\bar{\mathbf{p}}$ is the average power spectrum over all realizations. This covariance estimate can lead to a super-sampling effect \citep{Norberg2009, Takada2013, Lacasa2017, Lacasa2018}, since the generated patches overlap in some regions. This can lead to over- or underestimated errors in the likelihood analysis. We examine this effect in more detail in appendix \ref{ap:supersample} and found that it affects the area of our cosmological constraints in a random fashion by at most $\pm 5\%$.  Using the likelihood given by \cite{Selletin2016, Selletin2017} that incorporates the uncertainty of the covariance matrix  with flat priors $\Omega_m \in [0, 0.7]$ and $\sigma_8 \in [0.4, 1.4]$ one can then calculate the probability of measuring $\hat{\mathbf{p}}_\mathrm{meas.}$, if the true parameters are $\theta$, by calculating the conditional probability
\begin{equation}
P\left(\hat{\mathbf{p}}_\mathrm{meas.}\vert \theta, \hat{\Sigma}, N_s\right) \propto \left(1 + \frac{Q}{N_s - 1}\right)^{-\frac{N_s}{2}}, \label{eq:cons}
\end{equation}
where $Q$ is given by
\begin{equation}
Q = \left( \hat{\mathbf{p}}_\mathrm{meas.} - \mathbf{p}(\theta) \right)^T\hat{\Sigma}^{-1} \left( \hat{\mathbf{p}}_\mathrm{meas.} - \mathbf{p}(\theta) \right).
\end{equation}
One should also note that we did not assume a cosmology dependent covariance matrix and therefore absorbed the covariance dependent term of equation \eqref{eq:cons} into the normalization factor (compare to equation \eqref{eq:consCNN}). The same covariance matrix was used for all possible parameter configurations $\theta$. As mock observation $\hat{\mathbf{p}}_\mathrm{meas.}$ we chose the average power spectra of our fiducial cosmology, as it was done in \cite{Gupta2018}. This ensured that the resulting cosmological constraints were centered around our fiducial cosmology. The results of this analysis are presented in section \ref{sec:results}.

\section{Convolutional Neural Network \label{sec:CNN}}

We use the CNN to regress directly from convergence mass maps to cosmological parameter.
Then we build a likelihood described below and use it to calculate the parameters constrains.
In effect the network learns the optimal way to infer the parameters for the architecture used.

\subsection{Architecture}

The CNN operates on a single projected patch of size 256 x 256 pixels. This patch is passed through 5 convolutional layers, each with a stride of 2, and without any padding applied to the boundaries (which effectively yields to downsampling the input patch). The first convolutional layer used 16 square filters with a size of 7 x 7 pixels. This was followed by two layers with 32 and 64 filters with a size of 5 x 5 pixels and the last 2 convolutional layers had 128 and 256 filters of size 3 x 3 pixels. The output of the last convolutional layer was then flattened and the first fully connected layer projected this flattened output onto 1024 hidden neurons. This was then followed by a fully connected hidden layer with 256 neurons and the last fully connected layer outputted the final predictions. The architecture of the CNN is visualized in figure \ref{fig:architecture}. The total number of trainable parameters was roughly $10^7$. All convolutional and fully connected layers, except the final output layer, used a non-linear activation function, for which we chose the linear rectified unit
\begin{equation}
f(x) =
\left\{
\begin{array}{c}
x \qquad x \geq 0 \\
0 \qquad x < 0
\end{array}
\right.
.
\end{equation}
Similar to \cite{Levasseur2017}, we chose a negative log-likelihood loss as cost function
\begin{equation}
L = \frac{1}{2}\left(\ln\left(\left|\Sigma\right|\right) + \left( \theta_p - \theta_t \right)^T\Sigma^{-1}\left( \theta_p - \theta_t \right) \right), \label{eq:loss}
\end{equation}
where
\begin{equation}
\Sigma =
\begin{pmatrix}
\sigma_{\Omega_M}^2 & \mathrm{Cov}(\sigma_8, \Omega_M) \\
\mathrm{Cov}(\sigma_8, \Omega_M) & \sigma_{\sigma_8}^2
\end{pmatrix},
\end{equation}
is the covariance matrix containing the variances of the predicted parameters and their covariance, $\theta_p = (\Omega_M^p, \sigma_8^p)$ are the predicted values and $\theta_t = (\Omega_M^t, \sigma_8^t)$ are the labels. To use this cost function it was necessary for the network to output 5 values, which correspond to the parameters $\sigma_8$, $\Omega_M$, their variances $\sigma_{\Omega_M}^2$, $\sigma_{\sigma_8}^2$ and their covariance $\mathrm{Cov}(\sigma_8, \Omega_M)$. However, the CNN should always predict a valid positive semidefinite covariance matrix for any given input. Therefore, to increase the numerical stability \cite{Kendall2017} the network was not predicting the variances and the covariances directly, but rather $s_{\Omega_M} = \log\left( \sigma_{\Omega_M}^2 \right)$, $s_{\sigma_8} = \log\left( \sigma_{\sigma_8}^2\right)$ and $c = \tanh^{-1}\left( \mathrm{Corr}(\sigma_8, \Omega_M) \right)$, where
\begin{equation}
\mathrm{Corr}(\sigma_8, \Omega_M) = \frac{\mathrm{Cov}(\sigma_8, \Omega_M)}{\sigma_{\Omega_M}\sigma_{\sigma_8}} \in [-1, 1]
\end{equation}
is the correlation coefficient. This way the output range of the CNN was not restricted.

This cost function was chosen to capture the degeneracy between $\Omega_M$ and $\sigma_8$ when inferred from convergence maps. A priori it was not clear if the CNN would be able to optimize its predictions for $\Omega_M^p$ and $\sigma_8^p$ independently. This approach enabled the network to correlate its prediction for these two parameters and minimize its loss according to the overall prediction. Further we noticed a faster convergence of the network with the log-likelihood loss compared to the square loss function
\begin{equation}
L_\mathrm{square} = \left( \Omega_M^p - \Omega_M^t \right)^2 + \left( \sigma_8^p - \sigma_8^t \right)^2.
\end{equation}

One should note that we did not provide any labels for the variances and the covariance of the predicted cosmological parameters $\theta_p$. During the training the predictions of the variances and the cosmological parameters balanced each other in the following way. The first term only depends on the predicted covariance matrix. This term of the cost function ensures that the predictions of the cosmological parameters $\theta_p$ are close to their labels $\theta_t$ which also leads to small predicted variances. The second term couples the predicted covariance matrix and the predictions of the cosmological parameters $\theta_p$. It is large if the difference of the predicted cosmological parameters $\theta_p$ and their corresponding labels $\theta_t$ is large compared to their predicted variances. Optimizing this cost function should therefore lead to the prediction of cosmological parameters $\theta_p$ with minimal predicted variances.

\subsection{Training Strategy \label{sec:trainstrat}}

\begin{figure}
\includegraphics[width=0.5\textwidth]{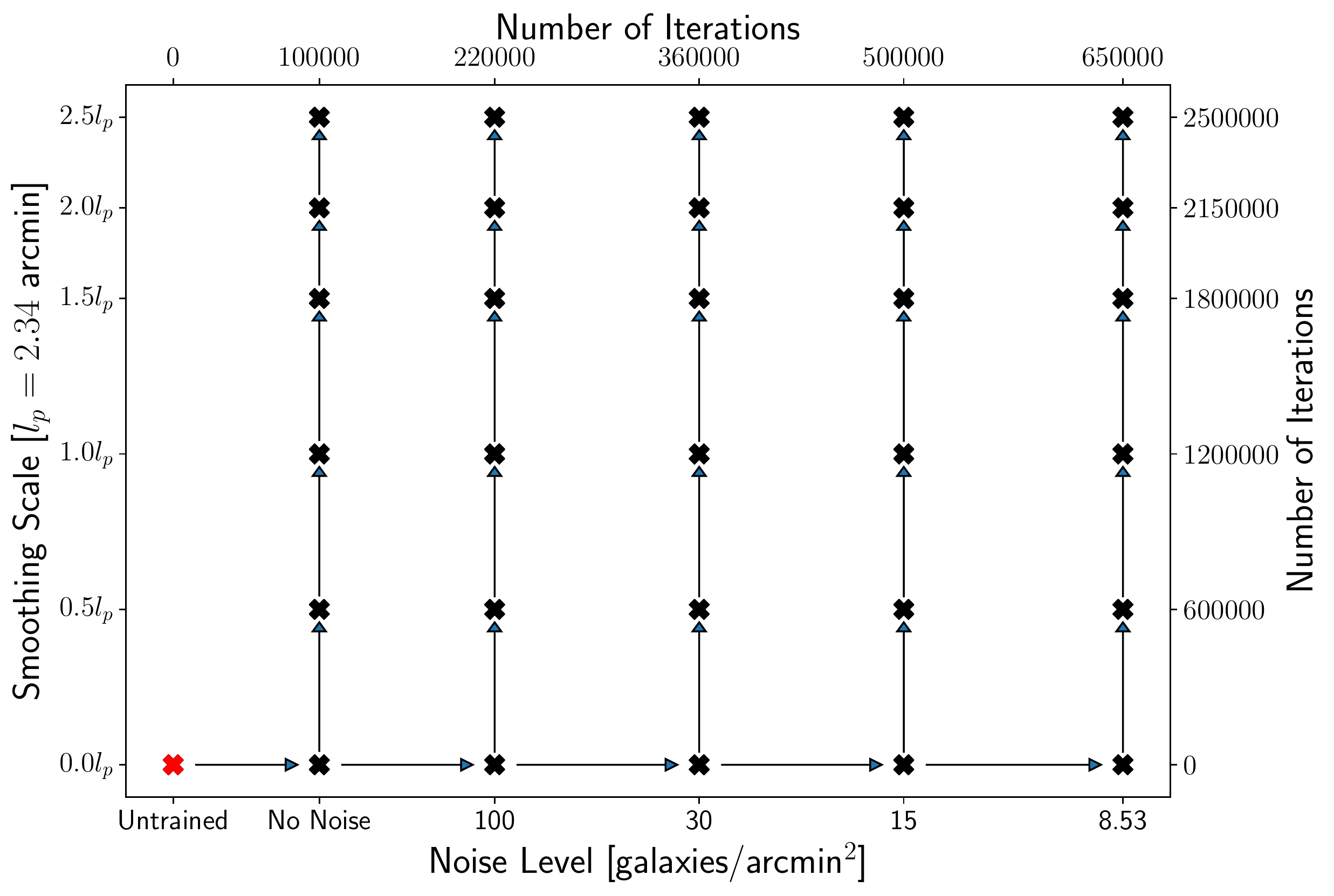}
\caption{Sketch of our training strategy. The labels of the axis on the left side and on the bottom show the corresponding smoothing scale and the noise level of different configurations, while the labels on the top and on the right side indicate the number of training iterations between two configuration. The red cross represents the untrained network and is the starting point of our training strategy. The black crosses show all our considered combinations of smoothing scales and noise levels. The arrows indicate the training of the CNN. \label{fig:trainstrat}}
\end{figure}

\begin{figure*}
\includegraphics[width=1.0\textwidth]{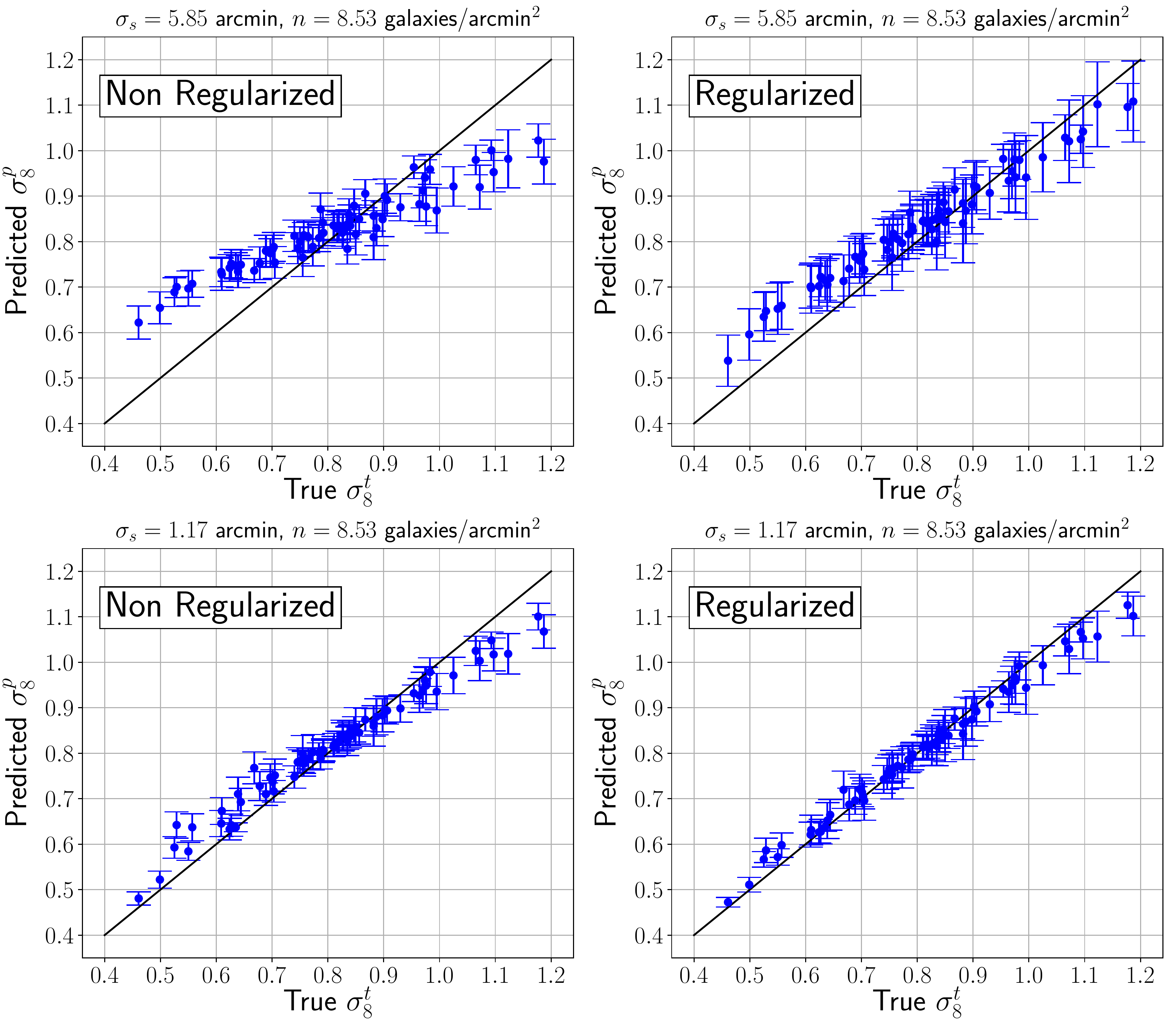}
\caption{Predictions of $\sigma_8^p$ of the CNN evaluated on the test set compared to their true values. The applied smoothing scale and the noise level can is written on top of each panel. The left panels show the predictions of the CNN before we applied the regularizing term (equation \eqref{eq:regu}) to our cost function (equation \eqref{eq:loss}). The right panels show the predictions of the network after the training with the regularizing term. Each blue dot corresponds to the mean prediction of the given cosmology and the errorbars indicate the standard deviation of all predictions. \label{fig:bias}}
\end{figure*}

Training the parameters of the CNN required splitting our dataset into two disjoint training and test sets. The training set consists of 600 patches from each full sky convergence map, resulting in a total of 2600 individual patches for each cosmology. We did not include additional data for our fiducial cosmology to prevent the CNN from becoming biased towards it. The test set consists of the remaining 168 patches per full sky convergence map, resulting in a total of 672 individual patches. The trainable parameters of the CNN were optimized using the \texttt{Adam} optimizer \cite{Kingma2014} with first and second moment exponential decay rates of 0.9 and 0.999. The initial learning rate was set to $10^{-4}$. The training set was split into batches of 32 patches. After one iteration through the whole training set, called an epoch, these batches were randomly reshuffled. Since the patches were already overlapping in some regions we did not use any form of input data augmentation like random rotations or random shifts. We will now detail the training procedure of the network. An important aspect of this procedure concerns the fact that the noise level and smoothing scale have to be adapted progressively in order to allow the network to learn to capture the relevant information in the input data. 

The whole training strategy illustrated in figure \ref{fig:trainstrat} consists of three parts and an example of the training and validation loss is shown in figure \ref{fig:trainloss} of appendix \ref{ap:loss}. In the first part we trained the network on unsmoothed images and increased the noise level. The network was first trained without any noise for 100'000 iterations. For the first 50'000 iterations we set the learning rate to $10^{-4}$ and to $2\times 10^{-5}$ for the next 50'000 iterations. One iterations corresponded to the optimization procedure over a batch of 32 patches. The learning rate was decreased to achieve a better convergence. Further it turned out that a learning rate of $10^{-4}$ was too high to increase the noise level. The reason for this is most likely that the gradient of the CNN became less stable while increasing the noise. By decreasing the learning rate the network will adjust it weights less after each iteration. The weights of this trained network were then saved to be reused in the later training stages. Afterwards the noise level was increased after each iteration. The increment of the noise was set to be $0.25\times 10^{-5}$ times the maximum considered noise level corresponding to a galaxy density of 8.53 galaxies/arcmin$^2$. After reaching a noise level corresponding to 100 galaxies/arcmin$^2$ we stopped increasing the noise and trained the network with a constant noise level for 50'000 iterations. The weights of the CNN were then saved again and the procedure was repeated until we reached our maximal examined noise level corresponding to a galaxy number-density 8.53 galaxies/arcmin$^2$.

In the second part of our training strategy we increased the applied smoothing scale. We started by linearly increasing the smoothing scale over 500'000 iterations up to a width of $\sigma_s = 0.5l_p$, where $l_p = 2.34$ arcmin was the length of a single pixel. We then trained the network for another 100'000 iterations with a fixed smoothing scale and saved the trainable parameters of the network. This procedure was then repeated up to a smoothing scale of $\sigma_s = 1.5l_p$. Afterwards we increased the smoothing scale increments. We linearly increased the smoothing scale up to $\sigma_s = 2.0l_p$ over 250'000 iterations and then trained the network for another 100'000 iterations with a constant smoothing scale. This was then repeated to increase the smoothing scale up to $\sigma_s = 2.5l_p$. We found that the network was extremely sensitive to a change in the smoothing scale for $\sigma_s \leq 1.5$. Increasing the smoothing scale too fast in this regime lead to a decrease in the networks performance. However, this sensitivity decreased for larger smoothing scales, allowing us to increase the smoothing scale increments. We also examined an exponential increase of the smoothing scale, which lead to similar results.

After the second part of our training strategy we let the CNN predict the parameters of the 168 realizations of our examined region in the test set. This was done by feeding in all individual patches of the test set into the CNN. The prediction of one realization was then calculated by averaging the predictions of the 4 individual patches in this realization. We noticed an obvious bias in the predictions of the network for high noise levels and large smoothing scales. An example of this bias for the parameter $\sigma_8$ is shown on the left panels of figure \ref{fig:bias}. The parameter $\Omega_M$ was affected similarly by the bias. To reduce this bias we decided to add a regularizing term to our cost function
\begin{equation}
L_2 = \lambda_1\left(\left(\Omega_M^w - \Omega_M^t \right)^2 + \left(\sigma_8^w - \sigma_8^t \right)^2\right), \label{eq:regu}
\end{equation}
where $\Omega_M^w$ and $\sigma_8^w$ are exponential weighted averages of the current predictions and the last 100 predictions of a given cosmology
\begin{equation}
\Omega_M^w = \sum_{i = 1}^{100}w_i\Omega_M^i, \quad w_i = \frac{\exp\left(\lambda_2i\right)}{N} , \quad N = \sum_{i = 1}^{100}w_i,
\end{equation}
where $\Omega_M^0 = \Omega_M^p$ is the current prediction and we set the decay factor $\lambda_2$ such that the current prediction is weighted with  $w_0 = 0.05$. The parameter $\lambda_1 = 5000$ was chosen in such a way that the regularizing term was roughly of the same order of magnitude as the original cost function in equation \eqref{eq:loss}. The third part of the training strategy was to train the saved networks of the second part further with this new cost function. For each combination of noise levels and smoothing scales the network was trained for another 100'000 iterations. Afterwards we reevaluated the CNN on the test set. An example of the resulting predictions can be seen in the right panels of figure \ref{fig:bias}. One can clearly see that the regularization acts as a bias-variance trade off term. The spread of the predictions increases, while the bias is reduced. One reason for the bias occurring in the first place is the non-uniform sampling grid (see figure \ref{fig:simugrid}) and the applied shape noise. If the network, due to the applied noise and smoothing scale, is not able to reliably distinguish the cosmologies it can reduce its loss by moving its prediction closer to the center of the grid. This is due to the higher density of grid points in this region. Predictions in this region have a lower expected loss since the batches of 32 patches of one optimization step were uniformly sampled from all possible cosmologies. Introducing a regularizing term like \eqref{eq:regu} is one possible way to reduce this bias. Another possible option would be to weight the loss of cosmologies closer to the edge of the grid more. However, we found that, without knowing the uncertainty of the cosmological parameters introduced by the shape noise and the smoothing, finding the optimal weights is a non-trivial task.

\subsection{Inference \label{sec:CNNinference}}

\begin{figure*}
\includegraphics[width=1.0\textwidth]{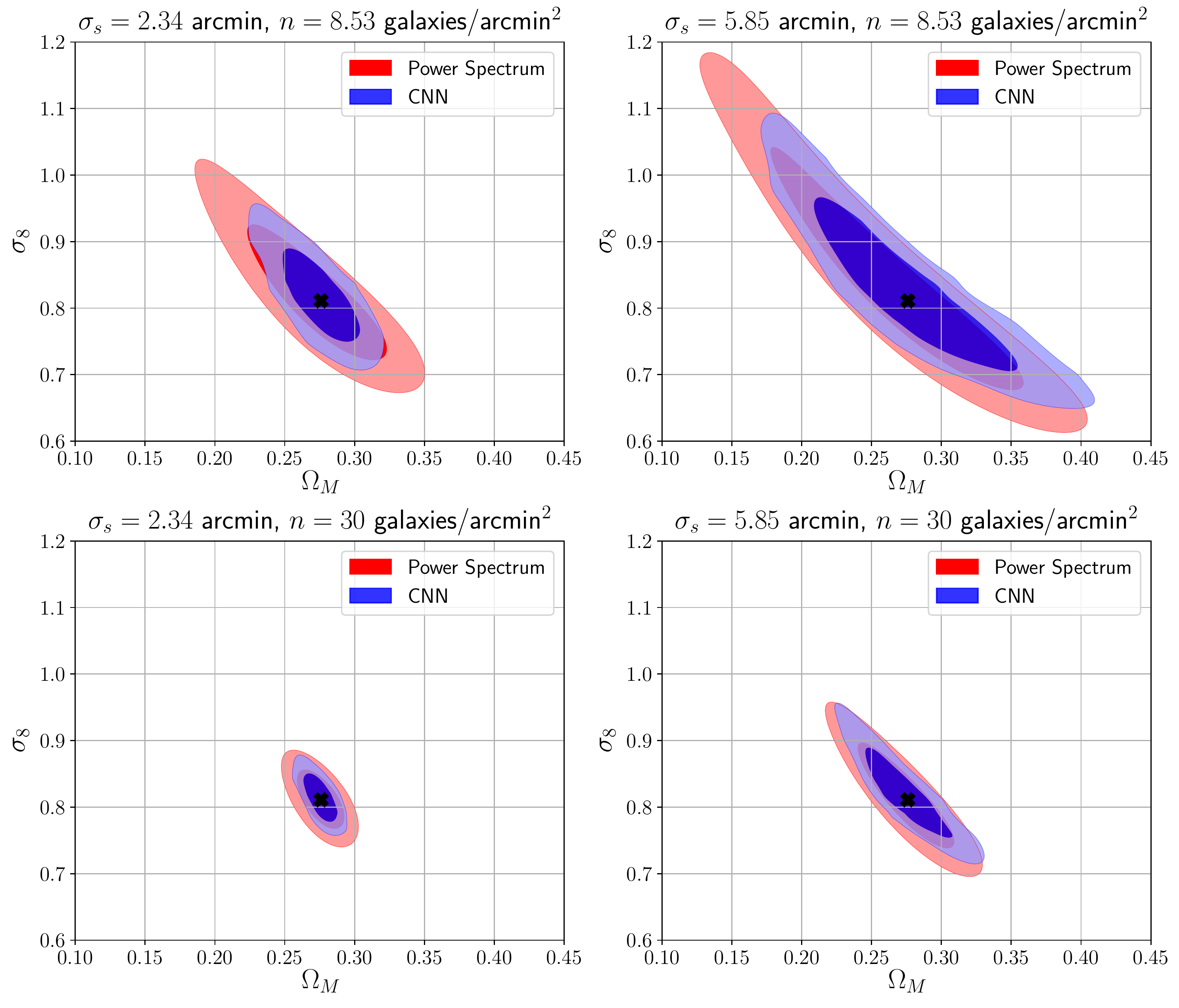}
\caption{Comparison of the cosmological constraints generated from the power spectrum analysis and the CNN. The noise level in two panels on the top corresponds to a galaxy number-density of $n = 8.53$ galaxies/arcmin$^2$. On the upper left panel we applied a Gaussian smoothing kernel of width $\sigma_s = l_p$ to the patches while on the upper right panel the width was set to $\sigma_s = 2.5l_p$, where $l_p = 2.34$ arcmin corresponds to the length of a single pixel of a patch. The black cross in the middle corresponds to our fiducial cosmology. In the lower two panels the noise level was set to $n = 30$ galaxies/arcmin$^2$ with the same smoothing scales as for the top panels.
 \label{fig:cons}}
\end{figure*}
\begin{figure*}
\includegraphics[width=1.0\textwidth]{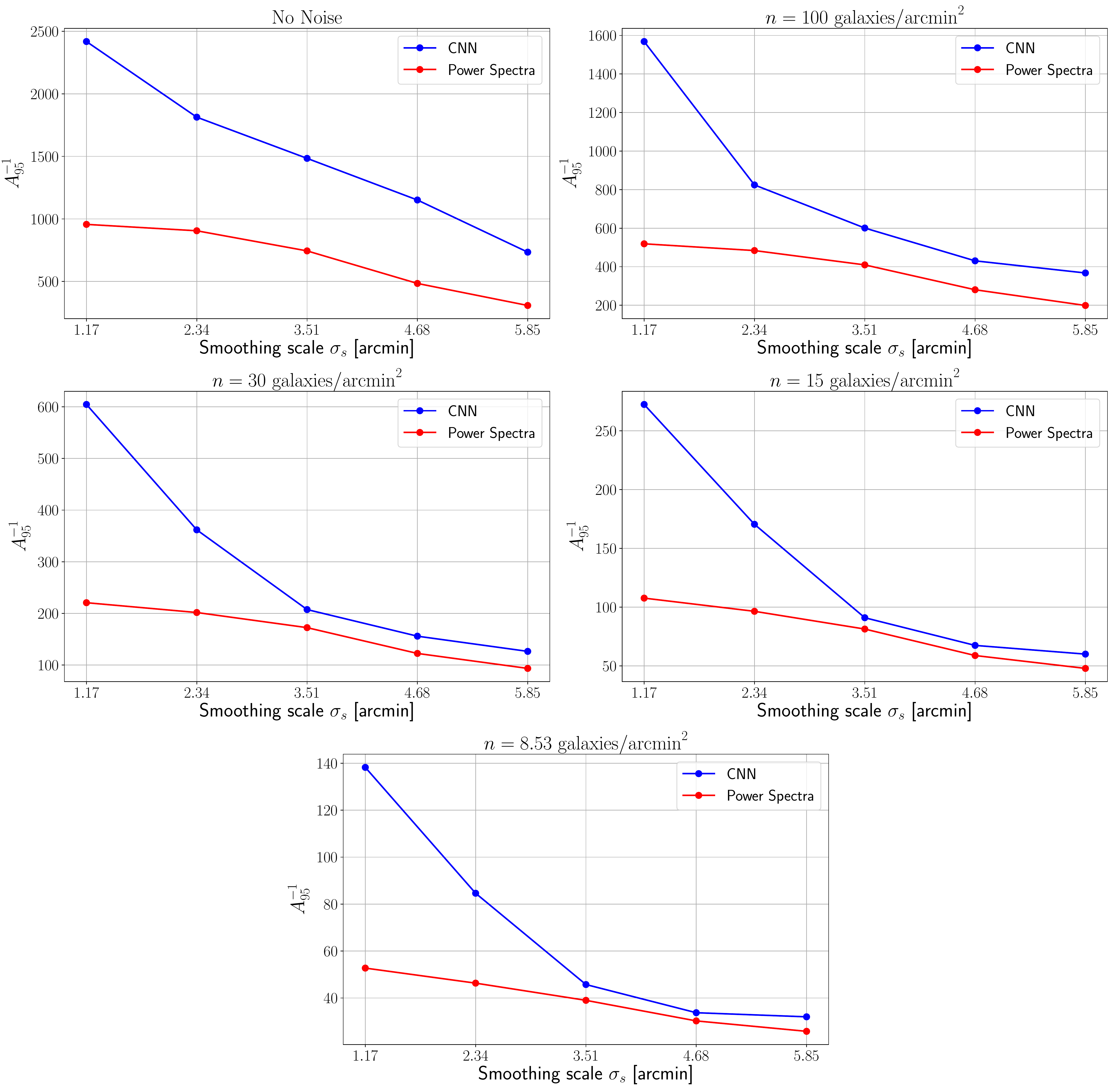}
\caption{Comparison of the cosmological constraints from the power spectrum analysis and the CNN. We chose the inverse of the area of the 95\% confidence contours in the $\Omega_M-\sigma_8$ plane $A_{95}^{-1}$ as figure of merit. The noise level is indicated by the galaxy number density on top of each panel. The horizontal axis corresponds to the width of the applied Gaussian smoothing kernel $\sigma_s$, which was always a multiple of the pixel length $l_p = 2.34$ arcmin.
 \label{fig:results}}
\end{figure*}
To generate cosmological constraints from the networks predictions we performed a similar Gaussian likelihood analysis as for the power spectrum. The network was evaluated on the 168 realizations of the examined region for each cosmology in the test set. We then calculated the average prediction $\bar{\mathbf{n}}_i$ of the network for each simulated cosmology. However, since the covariance matrix of these predictions depend strongly on the cosmology we decided to not only interpolate the means of these predictions, but also the covariance matrices. The effect on the cosmological constraints of an interpolated covariance matrix is examined in appendix \ref{ap:cov}. For numerical reasons we did not interpolate the covariance matrices directly but the parameters $s\left[\Omega^p_M\right] = \log( \sigma\left[\Omega^p_M\right]^2 )$, $s\left[\sigma^p_8\right] = \log( \sigma\left[\sigma^p_8\right]^2)$ and $c = \tanh^{-1}\left( \mathrm{Corr}(\sigma^p_8, \Omega^p_M) \right)$. Note that the variances and the covariance were calculated from the CNNs predictions of $\Omega^p_M$ and $\sigma^p_8$. The interpolation was done using unsmoothed linear radial basis functions. 

We did not use the network predictions of the covariance matrices for our main analysis. One reason was that our examined test region was consisting out of four independent patches and the input of the network was only one patch at a time. However, for single patches in a noise free setting and with our lowest considered smoothing scale, the network predictions of the covariance matrices matched the actual spread of its  predictions to a reasonable degree.

As for the power spectrum analysis, the constraints were then generated by calculating the conditional probability
\begin{equation}
P\left(\hat{\mathbf{n}}_\mathrm{meas.}\vert \theta, \hat{\Sigma}(\theta), N_s\right) \propto \left\vert \hat{\Sigma}_n(\theta) \right\vert^{-1/2}\left(1 + \frac{Q}{N_s - 1}\right)^{-\frac{N_s}{2}}, \label{eq:consCNN}
\end{equation}
with
\begin{equation}
Q = \left( \hat{\mathbf{n}}_\mathrm{meas.} - \mathbf{n}(\theta) \right)^T\hat{\Sigma}(\theta)^{-1} \left( \hat{\mathbf{n}}_\mathrm{meas.} - \mathbf{n}(\theta) \right),
\end{equation}
where $\hat{\mathbf{n}}_\mathrm{meas.}$ represents our mock observation, $\mathbf{n}(\theta)$ is the interpolated mean prediction of a given cosmology $\theta = (\Omega_M, \sigma_8)$ and $\hat{\Sigma}_n(\theta)$ the interpolated covariance matrix. One should note that the form of this likelihood differs slightly from the one in equation \eqref{eq:cons} because the interpolated covariance matrix can not be absorbed into the normalization factor. As for the power spectrum analysis, we chose as mock observation $\hat{\mathbf{n}}_\mathrm{meas.}$ the average prediction of the CNN of the 168 of the 400 deg$^2$ realizations of the convergence maps from our fiducial cosmology. Using this mock observation leads, by definition of the likelihood analysis, to constraints that are centered around our fiducial cosmology, which allows us to define contours of arbitrary size. That ensures a fair comparison to the power spectrum analysis. In appendix \ref{ap:robust} we explore the effect of the chosen mock observation on the cosmological constraints.

It is important to note that the bias of the network shown in figure \ref{fig:bias} does not affect the likelihood analysis. The CNN's prediction of our mock observation was not compared to the true parameters directly, but to the interpolated mean predictions of the network using Bayesian statistics. This way a bias in the predictions, as well as a higher variance, would lead to larger constraints. For example, a network that would output the parameters of our fiducial cosmology for every input would lead to flat posterior distribution spanning the whole area of our prior distribution, since equation \eqref{eq:consCNN} would lead to the same probability for every value of $\theta$.

\section{Results \label{sec:results}}

\begin{figure}
\includegraphics[width=0.5\textwidth]{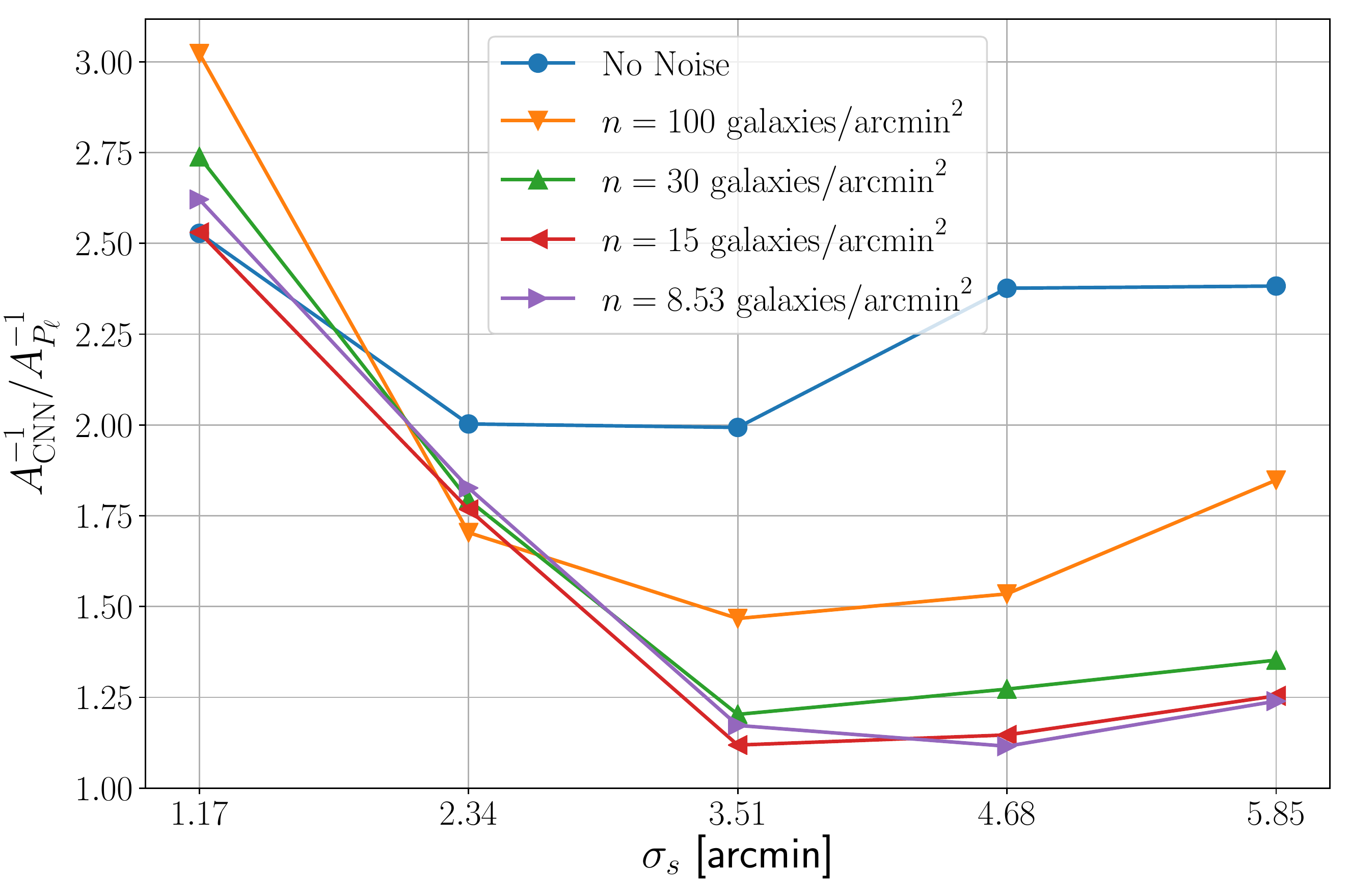}
\caption{The ratio of the figures of merit of the network $A^{-1}_\mathrm{CNN}$ and the power spectrum analysis $A^{-1}_{P_\ell}$ for all considered noise levels and smoothing scales. The applied Gaussian smoothing $\sigma_s$ was always an multiple of the pixel scale $l_p = 2.34$ arcmin.
 \label{fig:results_2}}
\end{figure}
As a figure of merit for the different methods we chose the inverse of the area of the 95\% confidence contours $A_{95}^{-1}$ in the $\Omega_M-\sigma_8$ plane. For constraints generated with a constant covariance matrix $\Sigma$ this figure of merit would be proportional to the inverse of the determinant $1/\det(\Sigma)$. An example of the cosmological constraints from two different settings is shown in figure \ref{fig:cons}. The top two panels of the figure show constraints where the noise level was set to $n = 8.53$ galaxies/arcmin$^2$, which corresponds to a realistic noise level for a ground-based observation. On the left panel we applied a Gaussian smoothing scale of width $\sigma_s = l_p$ to the patches while the smoothing scale on the right panel was set to $\sigma_s = 2.5l_p$. The constraints are perfectly centered around our fiducial cosmology since we chose the average results of our fiducial cosmology as mock observation. The constraints have roughly the same orientation, however, the CNN is able to generate tighter constraints. In the left panel the area of the 95\% confidence region generated by the CNN is 45\% smaller than the area of the confidence region of the standard power spectrum analysis and it is 19\% smaller in the right panel. The two panels on the bottom of the figure show constraints generated with a noise level corresponding to $n = 30$ galaxies/arcmin$^2$, which is a galaxy count that one can possibly achieve with a space-based observation. The constraints in the left panel were generated with an applied Gaussian smoothing scale of width $\sigma_s = l_p$ while the smoothing scale on the right panel was set to $\sigma_s = 2.5l_p$. The area of the 95\% confidence contours generated by the CNN are 44\% smaller in the left panel and 26\% smaller in the right panel compared to the ones from the standard power spectrum analysis.

The results from all considered settings are summarized in figures \ref{fig:results} and \ref{fig:results_2}. The panels of figure \ref{fig:results} show our figure of merit of the confidence contours generated from the CNN and the standard power spectrum analysis. One can clearly see that the CNN is outperforming the standard power spectrum analysis on all considered smoothing scales and all considered noise levels. However, the advantage of the CNN becomes smaller for higher smoothing scales and higher noise levels.

Figure \ref{fig:results_2} shows the ratio of the figures of merit of the network $A^{-1}_\mathrm{CNN}$ and the standard power spectrum analysis $A^{-1}_{P_\ell}$, which translates to the ratio of the area of their 95\% confidence contours $A^{P_\ell}_{95}$ and $A^\mathrm{CNN}_{95}$. For all smoothing scales and noise levels we see an improvement of the CNN over the power spectrum analysis. The advantage of the CNN decreases when the noise level and the smoothing scale is increased. Theoretically, one would expect that the confidence area of the CNN and the power spectrum analysis converge as the noise level and the smoothing scale increase. This is due to the reduced amount of information in such maps. Our results follow this trend.

For the lowest smoothing scale on all noise levels the ratio may be slightly overestimated because of non-vanishing power spectra modes $P_l$ above our considered range $\ell > 4000$.
This leads to a small advantage of the CNN since it can access more information than the power spectra. The next higher smoothing scale removes all modes above our considered range $P_l \sim 0$ for $\ell > 4000$ and the CNN and the power spectra have access to the same amount of information. For the noise free case in \cite{Gupta2018} the ratio of the confidence contours stays approximately constant for all smoothing scales. Our results follow that trend for smoothing scales $\sigma_s \geq l_p$. We therefore suspect that the advantage of the CNN for the lowest smoothing scale is overestimated by up to 20\%. For higher noise levels it is much harder to train the network and its relative advantage shrinks with respect to the applied noise level and smoothing scale. Further one can observe that on all considered noise levels the performance of the CNN is slightly worse than expected for a smoothing scale of $\sigma_s = 1.5l_p$. Such deviations from the general trend were also observed in \cite{Gupta2018}. This was most likely because of the chosen architecture of the network. When we experimented with different architectures we were able to improve the results for this specific smoothing scales but it also led to decreased performance in other regions.

\section{Conclusion and Outlook \label{sec:conc}}

Convolutional neural networks are potentially a powerful method to constrain cosmological parameters using weak lensing data.
In this work we were able to explore the advantages of a CNN compared to a standard power spectrum analysis in different survey regimes, by varying the noise level and smoothing scale.
We considered a 400 deg$^2$ survey scenario where the observed region was split into multiple patches.
We generated the convergence maps with a typical source galaxy redshift distribution and added a realistic shape noise.
We showed that the CNN is able to generate tighter constraints on all considered smoothing scales and noise levels.
However, the advantage of the CNN became smaller as the noise level and the smoothing scale increases.
For a noise level corresponding to a source galaxy number-density of $n = 8.53$ galaxies/arcmin$^2$ and an applied Gaussian smoothing kernel with a width of $\sigma_s = l_p$, where $l_p = 2.34$ arcmin was the length of a single pixel, the area of the 95\% confidence contours in the $\Omega_M - \sigma_8$ plane were 45\% smaller for the CNN compared to the standard power spectrum analysis.
We observe an even larger advantage of $\sim 50 \%$ for smaller smoothing scale of $\sigma_s = 1.17$.
This substantial improvement shows the great potential of analysing weak lensing maps with convolutional neural networks.

The application of CNNs to analyse real survey data will require careful consideration of systematic effects in the observed data, as well as of the accuracy and realism of the N-body simulations.
As CNNs can be sensitive to features in the data that were not present in the training set, the impact of these effect would have to be empirically quantified.
The effects of the errors in $n(z)$ estimation and in the shear calibration can be quantified by creating simulations that include these systematics \citep{Kacprzak2016a}.
The impact of astrophysical effects, such as Baryonic processes, intrinsic alignments, clustering of source galaxies, would most likely also have to be studied.
This may also be done by including them in simulations; a method was demonstrated to modify the N-body particles to emulate the effect of Baryons \citep{Schneider2015baryons}, while \citep{Joachimi2013intrinsic} developed a model for adding intrinsic alignments to the simulated data.
Furthermore, the impact of N-body simulation parameters, such as particle count, softening scale, or periodic box replication scheme, could be explored.

The robustness of the network to unknown systematics can also be studied in the framework of ``adversarial examples''~\cite{goodfellow2014explaining}, an active field of research in machine learning.
Using robust optimization techniques~\cite{ben2009robust} can help to make the network robust to a deterministic or set-based notion of uncertainty.

In this work we aimed to constrain only two cosmological parameters, $\Omega_M$ and $\sigma_8$.
Generation of the simulation data for larger parameter set may prove difficult, although simulation grids for extended cosmologies are becoming more common \citep{Heitman2014coyote,Lui2015peaks}.
Larger suites of simulations may be available in upcoming years, as the power of High Performance Computing machines grows.

Finally, it would be interesting to test the performance of the CNN for a tomographic analysis.
It may also be possible to include other maps, such as galaxy counts, cosmic microwave background temperature or lensing, to exploit the correlations between these probes \citep{des2017cosmology,vanUitert2017kidsgama,Nicola2016} in the framework of deep learning.



\begin{acknowledgments}
This work was supported by the Swiss Data Science Centre (SDSC), project \textit{sd01 - DLOC:  Deep Learning for Observational Cosmology}, and grant number 200021\_169130 from the Swiss National Science Foundation.
We thank Zoltan Haiman, Nathana\"{e}l Perraudin and Fernando Perez Cruz for helpful discussions.
\end{acknowledgments}

\appendix

\section{N-Body Simulation \label{ap:simulations}}

\begin{figure}
\includegraphics[width=0.5\textwidth]{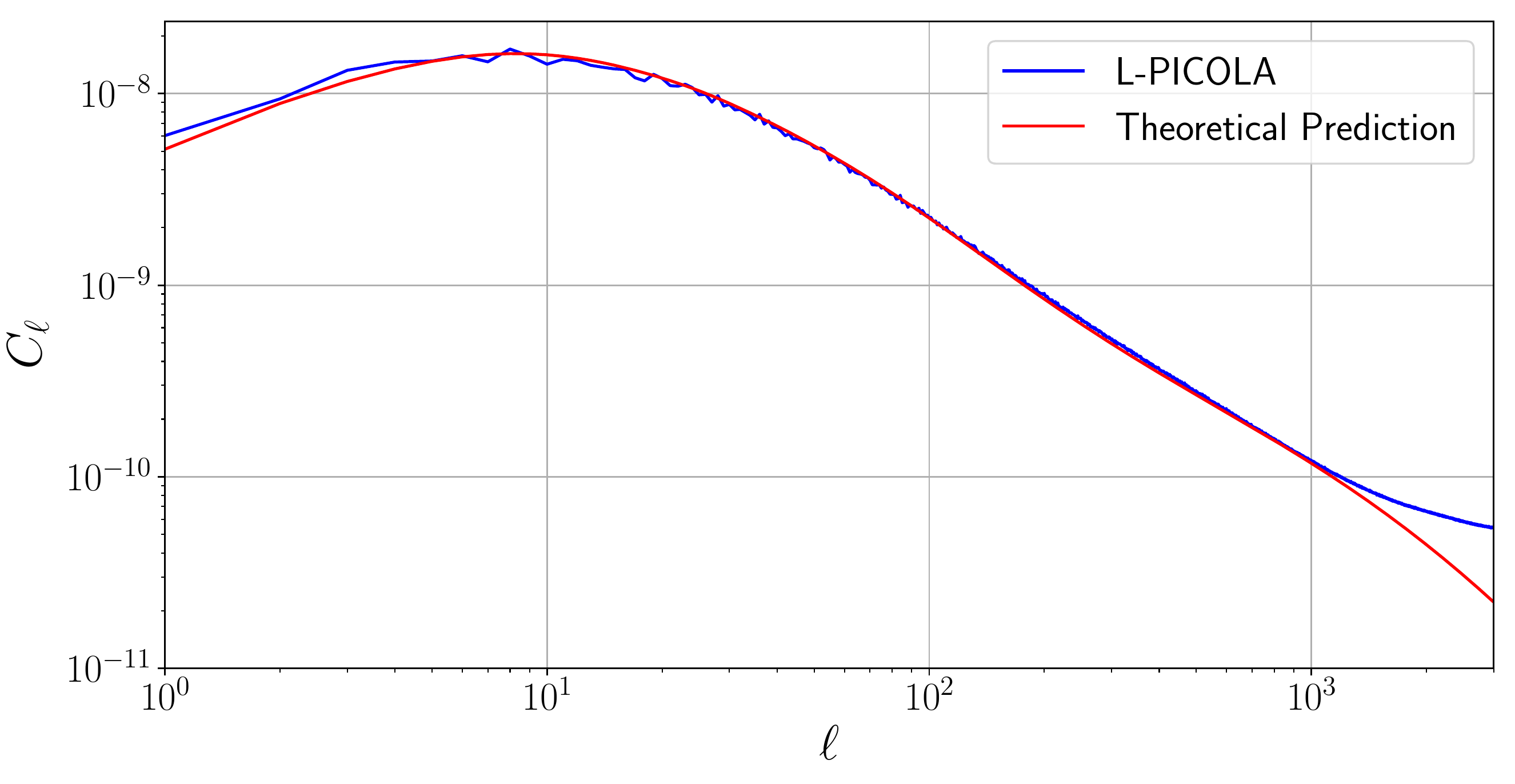}
\caption{Average power spectrum of noise free, full sky convergence maps generated from the 40 simulations of our fiducial cosmology. The theoretical prediction was calculated using \texttt{PyCosmo} \cite{Refregier2017}. \label{fig:power_spec}}
\end{figure}

Besides the number of mesh points, the simulation parameters were the same as in \cite{Fluri2018}. The reason we reduced the number of mesh points was to reduce the amount of computation time. This was necessary since the total number of simulations in this work was 4 times larger than in \cite{Fluri2018}. As suggested in \cite{Howlett2015} we started all our \texttt{L-PICOLA} simulations at redshift $z = 9$. For each generated full sky convergence map we simulated two independent boxes using the nested box approach. The fist box had $256^3$ particles, $512^3$ mesh points and a size of 1.5 Gpc. The second box had $256^3$ particles, $512^3$ mesh points and a size of 2.25 Gpc. The in section \ref{sec:data} mentioned 6 Gpc and 9 Gpc boxes were then generated using periodic boxes. Each of the bigger boxes was generated using 64 periodic boxes. To reduce the special correlations we applied random shifts and 90$^\circ$ rotations and parity flips to each of the smaller boxes. For the 6 Gpc box we used 10 time steps from $z = 9$ to $z = 0.8$, after that we used time steps of size $\Delta z = 0.01$ down to $z = 0.1$ and snapshots were outputted for each time step. Similarly for the 9 Gpc box we used 10 time steps from $z = 9$ to $z = 1.5$, after that we used time steps of size $\Delta z = 0.01$ down to $z = 0.8$ and again snapshots were outputted for each time step. Out of each snapshot we cut out a shell of thickness $\Delta z = 0.01$ to generate a past lightcone which mimics the universe as we observe it. These shells were then used to generate full sky convergence maps using \texttt{Ufalcon} \cite{Sgier2018}. To test the accuracy of the generated simulation we calculated the average convergence power spectrum from the 40 simulations of our fiducial cosmology. The comparison of this average power spectrum with a theoretical prediction is shown in figure \ref{fig:power_spec}. The power spectrum agrees well up to $\ell \sim 1000$ which is similar to the results of \cite{Sgier2018}. The discrepancy of the power spectrum for $\ell > 1000$ arises mostly from the shot noise generated by the finite amount of particles in the simulations and projection effects, we expect it to be independent of the cosmology. Similar to \cite{Gupta2018}, we therefore included also modes $\ell > 1000$ in our analysis.
\begin{figure*}
\includegraphics[width=1.0\textwidth]{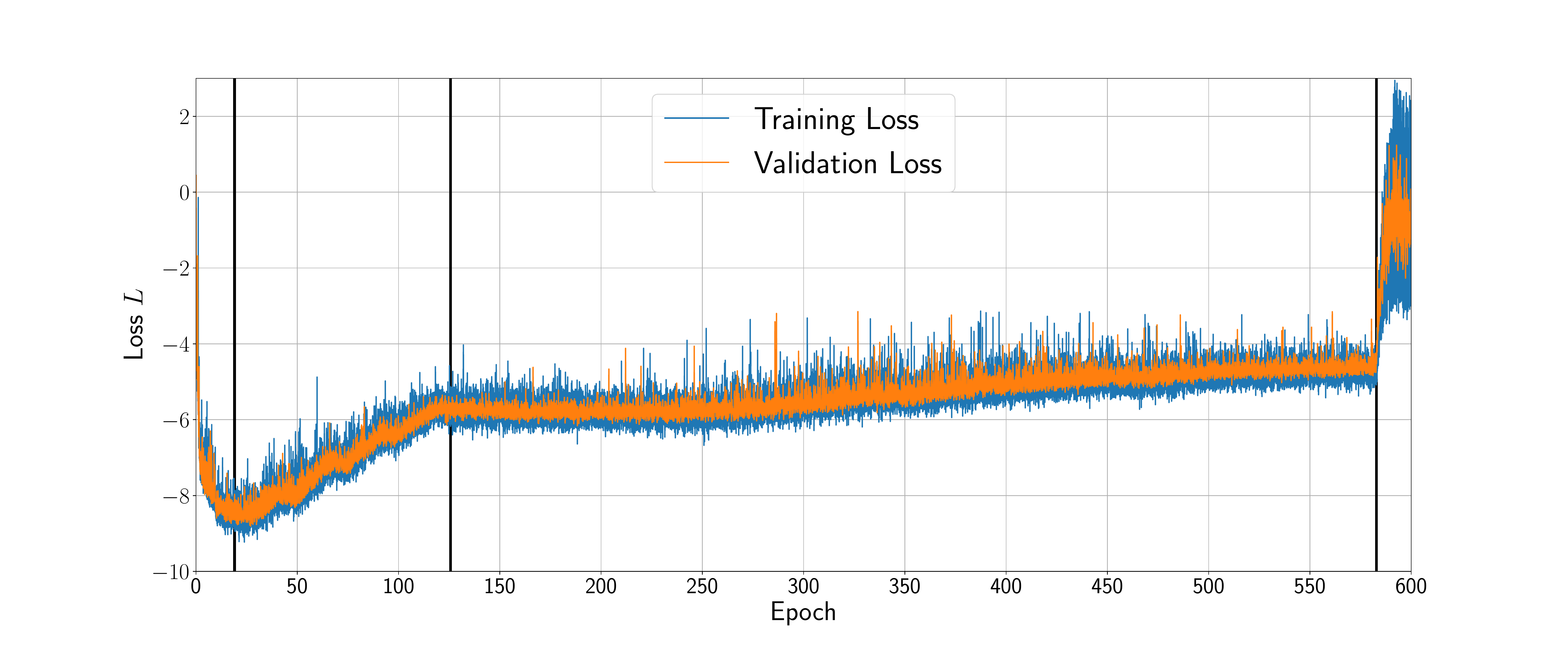}
\caption{The training and validation loss of the network. One epoch corresponds to one iteration through the whole training set. The black solid lines indicate different stages of the training (compare with figure \ref{fig:trainstrat}). The network starts with randomly initialized weights and up to the first solid line the training was performed with unsmoothed noise free convergence maps. Afterwards we started to increase the noise as described in section \ref{sec:trainstrat}. The maximum noise level corresponding to $n = 8.53$ galaxies/arcmin$^2$ was reached by the second solid line. Following the noise increments we started to increase to smoothing scale. The highest smoothing scale of $\sigma_s = 5.85$ arcmin was reached by the last solid line. The remaining training was then performed with the additional regulizer given by equation \ref{eq:regu}.   \label{fig:trainloss}}
\end{figure*}
\begin{figure}
\includegraphics[width=0.5\textwidth]{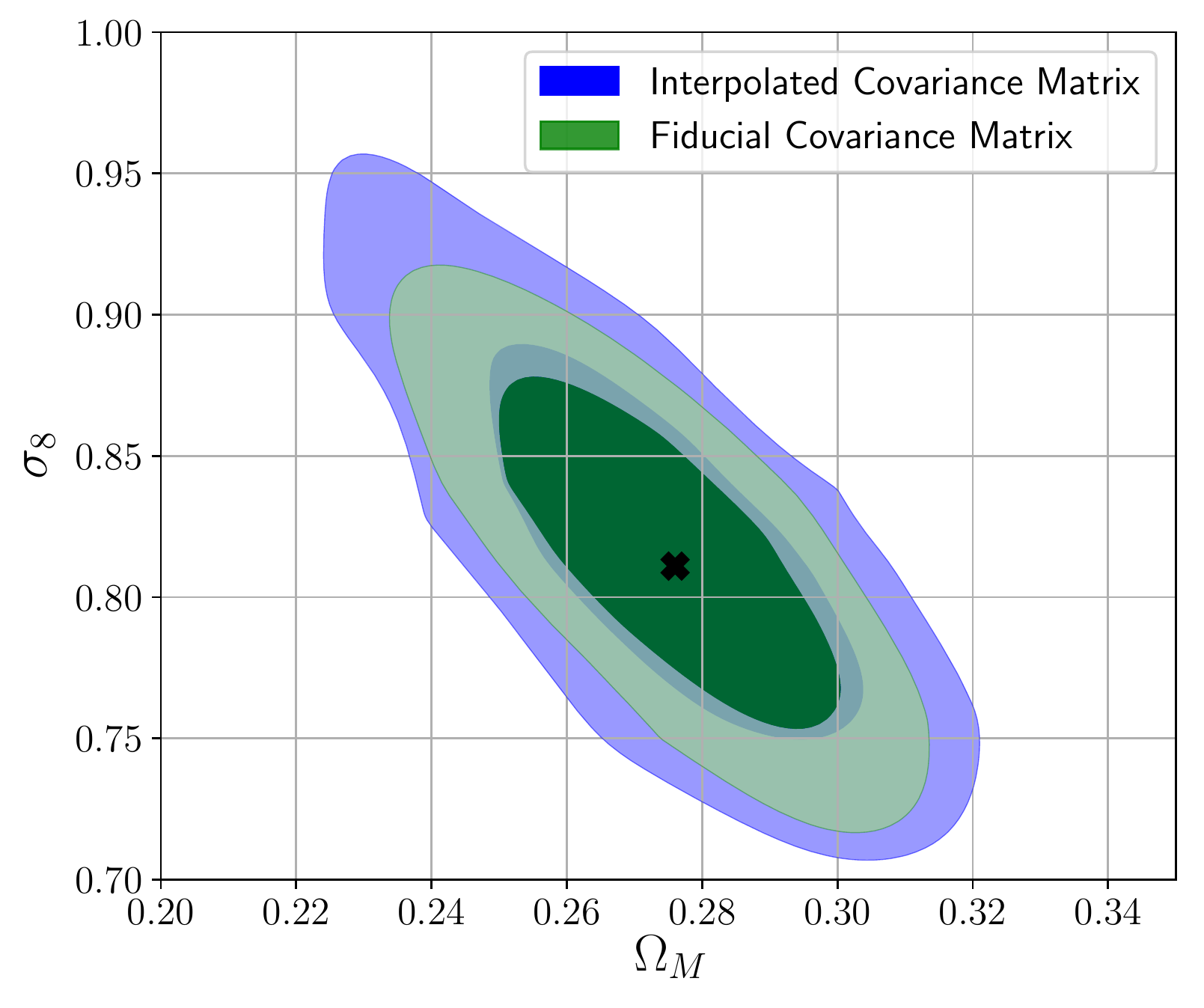}
\caption{The 68\% and 95\% confidence contours of cosmological constraints generated from the CNN. The green contours correspond to constraints where we only used the covariance matrix of our fiducial cosmology in the likelihood analysis. For the blue contours we interpolated the variances and the covariance of the predicted parameters through the grid. \label{fig:covcons}}
\end{figure}

\section{Full sky Convergence Map generation \label{ap:convmaps}}

The full sky convergence maps were generated using \texttt{Ufalcon} \cite{Sgier2018}. A more detailed approach can be found in \cite{Sgier2018}. \texttt{Ufalcon} \cite{Sgier2018} uses the Born approximation. We expect the Born approximation to be valid for our chosen pixel size of $l_p = 2.34$ arcmin. However, full ray-tracing may be necessary to correctly resolve smaller structures \cite{Petri2017Born}. With the connection between the convergence and the overdensity, the convergence at a given pixel $\theta_\mathrm{pix}$ can be calculated using
\begin{equation}
\kappa(\theta_\mathrm{pix}) \approx \frac{3}{2}\Omega_\mathrm{m}\sum_bW_b\frac{H_0}{c}\int_{\Delta z_b}\frac{c\mathrm{d}z}{H_0E(z)}\delta\left(\frac{c}{H_0}\mathcal{D}(z)\hat{n}_\mathrm{pix},z\right),
\end{equation}
where $\mathcal{D}(z)$ is the dimensionless comoving distance, $\hat{n}_\mathrm{pix}$ is a unit vector pointing to the pixels center and $E(z)$ is given by
\begin{equation}
\mathrm{d}\mathcal{D} = \frac{\mathrm{d}z}{E(z)}.
\end{equation}
The sum runs over all redshift shells and $\Delta z_b = 0.01$ is the thickness of the shell. Each shell gets the additional weight $W_b$ which depends on the redshift distribution of the source galaxies. For a given source redshift distribution the weight can be calculated using
\begin{equation}
W^{n(z)}_b = \frac{\left(\int_{\Delta z_b}\frac{\mathrm{d}z}{E(z)}\int_z^{z_s}\mathrm{d}z'n(z')\frac{\mathcal{D}(z)\mathcal{D}(z,z')}{\mathcal{D}(z')}\frac{1}{a(z)}\right)}{\left(\int_{\Delta z_b}\frac{\mathrm{d}z}{E(z)}\int_{z_0}^{z_s}\mathrm{d}z'n(z')\right)},
\end{equation}
where $z_0$ is the redshift of the first shell and $z_s$ the redshift of the last shell that is added. In this work we always used the redshift boundaries $z_0 = 0.1$ and $z_s = 1.5$.

\begin{figure*}
\includegraphics[width=1.0\textwidth]{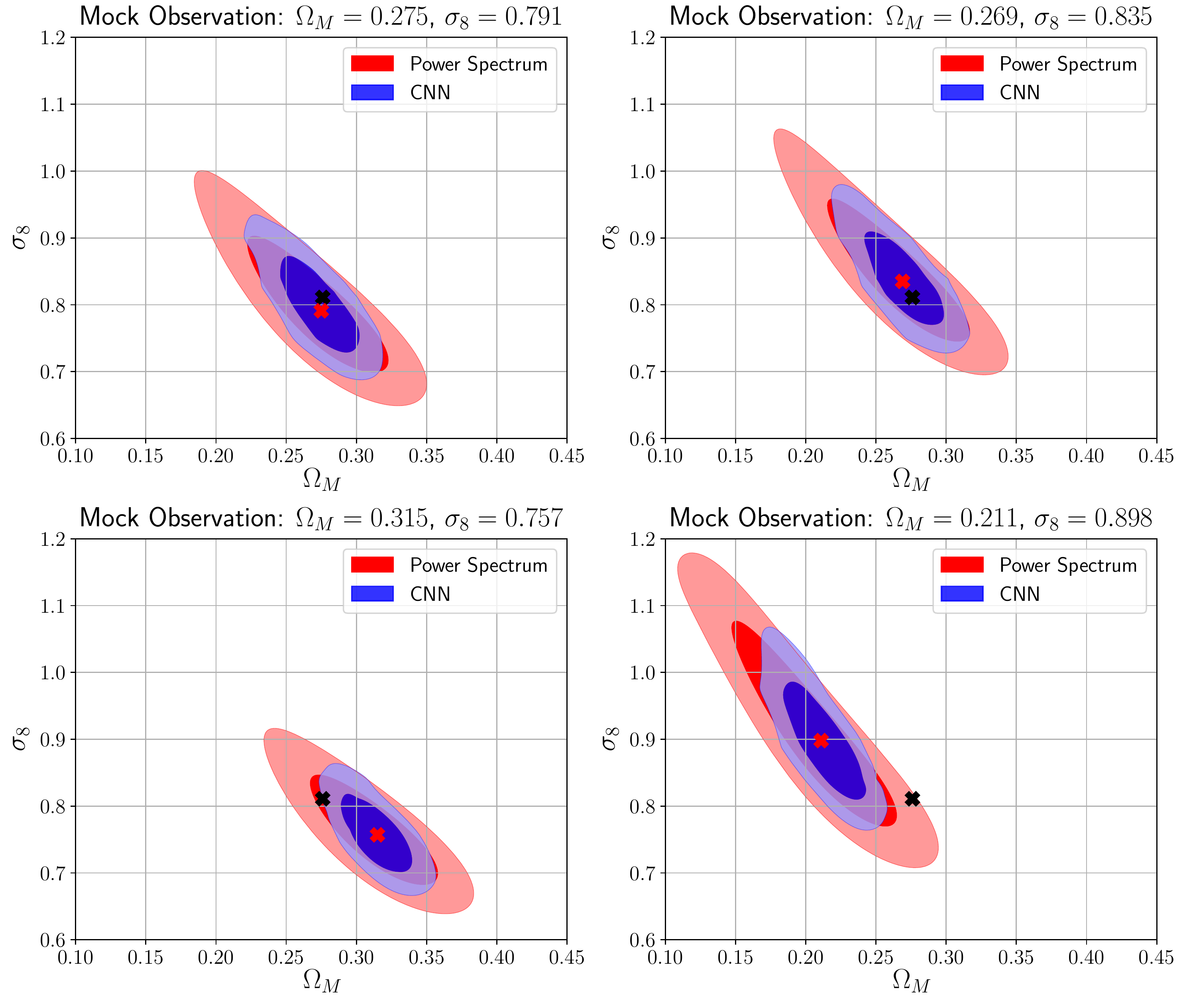}
\caption{Comparison of the cosmological constraints generated from the power spectrum analysis and the CNN for different mock observations. The noise level and the smoothing scale were set to $n = 8.53$ galaxies/arcmin$^2$ and $\sigma_s = 2.34$ arcmin in all panels. The cosmology used as mock observation is indicated on top of each panel and shown in the plots as red crosses. The black crosses show our original fiducial cosmology.
 \label{fig:robust}}
\end{figure*}
\section{Super-Sample Covariance \label{ap:supersample}}

Using the same information multiple times in a covariance estimate can cause a super-sampling effect \citep{Norberg2009, Takada2013, Lacasa2017, Lacasa2018} and can lead to over- or underestimated errors in the likelihood analysis.
The convergence maps used to estimate the covariance matrices were overlapping in some regions (see Section \ref{sec:powerinfer}).
These overlapping regions, however, had different rotations and realisations of noise, so they were not fully dependent.

To estimate the scale of this effect on the generated constraints we performed a second likelihood analysis where we used only non-overlapping realizations of our examined region.
For the power spectrum we used 1460 realizations of our fiducial cosmology to estimate the covariance matrix.
We noticed that, on all considered noise levels and smoothing scales, the area of the 95\% confidence contours was changing by $\pm 5\%$, with the mean shift $<1\%$.
For the network, we used 89 non-overlapping realizations of our examined region from our test set to estimate the covariance matrices.
A similar change of the confidence contours was observed; the contours were varying by $\pm 5\%$.
The mean, however, increased by rougly $1-2\%$.
This is mainly due to the fact that the uncertainty of the covariance increased.
It had more pronounced effect on the CNN, where the number of maps changed from 168 to 89, while the spectrum was left with 1460, which is already a sufficient number for covariance calculation for 19 bins.

One should note that the reduced number of realizations also lead to slight changes in the mean spectra and mean predictions of the network for our considered cosmologies.
Further, it also influenced our mock observations, since we chose them to be the mean prediction or power spectra of our fiducial cosmology.
These effects, besides the super-sampling covariance, can also lead to small changes in the results.

The observed change in the results did not depend on the applied noise level.
We would expect the super-sampling effect to be smaller for higher noise levels, since the overall information in the overlapping regions is reduced.
Since removing the overlapping regions lead to a change independent of the noise level, we suspect the super-sampling effect to be small.

The periodic boxes used in the lightcone generation could potentially also affect the covariance estimate. However, the results from Petri \cite{Petri2016Hai} indicate that this effect is not dominant.

Given that the difference in contour area between using 89 and 168 patches for covariance estimation is small, and that the overlapping realisations are not fully dependent (same structures with different rotation and noise realisation), we condlude that super-sampling effects do not significantly affect our results.

\section{Training and Validation Loss \label{ap:loss}}

One example of the training and validation loss of our network is shown figure \ref{fig:trainloss}. It shows the training and validation loss up to the highest considered noise level and the highest considered smoothing scale. It can be seen that the training and validation loss are almost equivalent throughout the training. One can therefore assume that the network is not overfitting. The validation loss was evaluated on 256 randomly selected convergence maps of the test set every 250 iterations. The larger batches of the validation loss explain its smaller spread compared to the training loss, which was evaluated on 32 convergence maps.

\section{Covariance Matrix Interpolation \label{ap:cov}}

The variances and the covariance of the CNN's prediction of the cosmological parameter $\theta_p = (\Omega_M^p, \sigma_8^p)$ depends strongly on the cosmology. This can for example be seen in figure \ref{fig:bias}. Using the same same covariance matrix for all cosmological parameters as we did for the power spectrum analysis can therefore lead to underestimated cosmological constraints. Figure \ref{fig:covcons} shows the cosmological constraints generated from the CNN where we applied a smoothing scale of $\sigma_s = l_p$ to the patches and the noise level was set to $n = 8.53$ galaxies/arcmin$^2$. The blue contours correspond to a likelihood analysis where we only used the covariance matrix of our fiducial cosmology. For the red contours we interpolated the covariance matrix as explained in section \ref{sec:CNNinference}. One can clearly see that interpolating the covariance matrix leads to larger constraints. The area of the 95\% confidence contours increases by 47\%.

\section{Robustness of Constraints \label{ap:robust}}

In a realistic scenario it is not possible to know the cosmological parameters of the observation beforehand. It was therefore important to test the robustness of our generated constraints dependent on the mock observation. To do this we generated cosmological constraints using the CNN and the standard power spectrum analysis for different mock observations. The resulting constraints are shown in figure \ref{fig:robust}. To generate these constraints we repeated the same likelihood analysis as for our fiducial cosmology, but replacing the mock observation with the mean prediction or mean power spectra of different cosmologies. Figure \ref{fig:robust} clearly shows that the constraints are robust under such a change. The area of the 95\% confidence contours generated from the CNN are $\sim 50\%$ smaller than the constraints generated from the standard power spectrum analysis for all shown mock observations.

\bibliography{library}

\end{document}